\documentclass[manuscript]{aastex}

\newcommand{\brg}{$Br\,\gamma$ }
\newcommand{\pfa}{$Pf_{\rm avg}$ }
\newcommand{\dph}{$d\phi$ }
\newcommand{\kms}{${\rm km\,s}^{-1}$ }
\newcommand{\pcs}{{\it PC }$_{\rm line}^{\rm shift}$~}
\newcommand{\thl}{$\theta_{\rm line}^{\rm FWHM}$}

\shorttitle{Spectro-interferometry of 48~Lib}
\shortauthors{Pott et al.}

\begin{document}

\title{Probing local density inhomogeneities in the circumstellar disk of a Be star using the new spectro-astrometry mode at the Keck interferometer}


\author{J.-U. Pott\altaffilmark{1,2,3}, J. Woillez\altaffilmark{2}, S. Ragland\altaffilmark{2}, P. L. Wizinowich\altaffilmark{2}, J. A. Eisner\altaffilmark{4}, J. D. Monnier\altaffilmark{5},
R. L. Akeson\altaffilmark{6}, A. M. Ghez\altaffilmark{3,7}, J. R. Graham\altaffilmark{8}, L. A. Hillenbrand\altaffilmark{9}, R. Millan-Gabet\altaffilmark{6}, E. Appleby\altaffilmark{2}, B. Berkey\altaffilmark{2}, M. M. Colavita\altaffilmark{10}, A. Cooper\altaffilmark{2}, C. Felizardo\altaffilmark{6},
J. Herstein\altaffilmark{6}, M. Hrynevych\altaffilmark{2}, D. Medeiros\altaffilmark{2}, D. Morrison\altaffilmark{2}, T. Panteleeva\altaffilmark{2}, 
B. Smith\altaffilmark{2}, K. Summers\altaffilmark{2}, K. Tsubota\altaffilmark{2}, C. Tyau\altaffilmark{2}, E. Wetherell\altaffilmark{2}}

\altaffiltext{1}{Max-Planck-Institut f{\" u}r Astronomie, K{\" o}nigstuhl 17, D-69117 Heidelberg, Germany \email{jpott@mpia.de}}
\altaffiltext{2}{W.~M.~Keck~Observatory, California Association for Research in Astronomy, Kamuela, HI 96743}
\altaffiltext{3}{Div. of Astronomy \& Astrophysics, University of California, Los Angeles,  CA 90095-1547}
\altaffiltext{4}{Steward Observatory, University of Arizona, Tucson, AZ 85721}
\altaffiltext{5}{Astronomy Department, University of Michigan, Ann Arbor, MI 48109}
\altaffiltext{6}{NASA Exoplanet Science Institute, Caltech, Pasadena, CA 91125}
\altaffiltext{7}{Institute of Geophysics and Planetary Physics, University of California, Los Angeles, CA 90095-1565}
\altaffiltext{8}{Astronomy Department, University of California Berkeley, CA 94720, USA}
\altaffiltext{9}{California Institute of Technology, Pasadena, CA 91125, USA}
\altaffiltext{10}{Jet Propulsion Laboratory, California Institute of Technology, Pasadena, CA
91109, USA}

%



\begin{abstract}
We report on the successful science verification phase of a new observing mode at the Keck interferometer, which provides a line-spread function width and sampling of 150~km/s at $K'$-band, at a current limiting magnitude of $K'\,\sim\,7$~mag with spatial resolution of $\lambda\,/\,2\,B\,\approx\,2.7\,{\rm mas}$ and a measured differential phase stability of unprecedented precision (3~mrad at $K\,=\,5\,{\rm mag}$, which represents 3~$\mu$as on sky or a centroiding precision of $10^{-3}$). 
The scientific potential of this mode is demonstrated by the presented observations of the circumstellar disk of the evolved Be-star 48~Lib. 
In addition to indirect methods such as multi-wavelength spectroscopy and polaritmetry,  the here described spectro-interferometric astrometry provides a new tool to directly constrain the radial density structure in the disk. 
We resolve for the first time several Pfund emission lines, in addition to \brg, in a single interferometric spectrum, and with adequate spatial and spectral resolution and precision to analyze the radial disk structure in 48~Lib. 
The data suggest that the continuum and $Pf$-emission originates in significantly more compact regions, inside of the \brg emission zone. 
Thus, spectro-interferometric astrometry opens the opportunity to directly connect the different observed line profiles of \brg and Pfund in the total and correlated flux to different disk radii.
The gravitational potential of a rotationally flattened Be star is expected to induce a one-armed density perturbation in the circumstellar disk. 
Such a slowly rotating disk oscillation has been used to explain the well known periodic V/R spectral profile variability in these stars, as well as the observed V/R cycle phase shifts between different disk emission lines. 
The differential line properties and linear constraints set by our data are consistent with theoretical models and lend direct support to the existence of a radius-dependent disk density perturbation. The data also shows decreasing gas rotation velocities at increasing stello-centric radii as expected for Keplerian disk rotation, assumed by those models. 
 
\end{abstract}



\section{Introduction}
\label{sec:1}
The ASTRA (ASTrometric and phase-Referencing Astronomy) upgrade program, funded by the NSF Major Research Instrumentation program, aims at extending the sensitivity and spectral resolution of the Keck interferometer (KI) through phase referencing and to implement a narrow-angle astrometry mode to enable a much broader scientific use of the KI \citep[][]{1992A&A...262..353S, 2007Wiz, 2008arXiv0811.2264P, 2010Woi}. 
In this paper we report on results of the successful implementation of the first ASTRA mode, the self phase referencing (SPR) mode.
In the SPR mode, 55~\% of the light from the on-axis science target is used to provide fringe tracking (i.e., phase referencing) information while 20~\% is used to make the science measurement.
Due to the stabilized fringes, $>$100x longer integrations can be taken on the second camera. 
Therefore, it is possible in SPR to increase the spectral resolution of the interferometric measurement by at least an order of magnitude. SPR as currently implemented and provided to the general user, offers 
a spectral resolution $R\,(\,=\,\lambda/\Delta \lambda)$ of 2000, which is Nyquist sampled at R = 1000 (330 pixels across the K'-band) at K $\lesssim\,7\,{\rm mag}$. 
Further technical aspects and results from the commissioning are discussed by \citet{2010Woi}. 
Besides simple amplitude spectro-interferometry at this resolution, the here presented data demonstrate that the differential visibility phase can be retrieved at a precision of at least 3~mrad for a bright star.
Non-linear phase changes in the source spectrum can be measured, and typically indicate translations of the photo-center over the respective spectral channels on the sky \citep[e.g.][]{2007A&A...464...87W}. 
The operational readiness of KI-SPR was successfully demonstrated with the here discussed observations of \objectname[48 Lib]{48~Lib} on the night of April 25, 2008 (UT). In addition, several young stellar objects were observed successfully \citep{2010Eis}. We show at the example of 48~Lib the wealth of spatial and spectral information that can be provided by SPR data. 

48~Lib is a well studied classical Be-star with circumstellar shell-emission lines.
Previous optical-nearinfrared long-baseline interferometric (OLBI) measurements of Be-star-shells demonstrated that 3d-models for H-line and continuum emission often overpredict the true size, which demonstrates the need for direct OLBI measurements \citep{2001A&A...367..532S,2005A&A...435..275C}. Narrow-band observations with the Mark III interferometer \citep{1988A&A...193..357S}, resulted in Be-star disk diameter estimates ranging in 2.6 to 4.5 mas \citep{1997ApJ...479..477Q}, well suited for the angular resolution of the 85 m baseline of the KI ($\lambda/2B\,=\,2.7~{\rm mas}$).
We chose 48~Lib as science demonstration target because recent detection of very narrow satellite absorption features in Fe~II shell lines reliably indicate very high inclination angles \citep{1996A&A...312L..17H}. This ensures strong features in the differential visibility and phase signals across the emission lines.

A large fraction of Be stars show cyclic variations in the ratio between the violet and red flux of the HI emission lines (V/R), an enigmatic phenomenon studied since almost a century \citep{1925PA.....33..537C}.  
Early measurements with the GI2T interferometer spatially and timely resolved the $H\,\alpha$ emission of $\gamma$-Cas, showing spatial variation of the line emission region with time \citet{1989Natur.342..520M}.
The fast rotation of Be-stars \citep[near-critical rotational velocities for cooler stars like 48~Lib,][]{2005ApJ...634..585C} results in a flattening of the stellar shape, which can be observationally confirmed by OLBI \citep{2003A&A...407L..47D}. 
While the rapid rotation might play a role in the ejection of the circumstellar envelope, the 
quadrupole moment of the gravitational potential of a flattened Be star is expected to create a prograde one-armed spiral density pattern, slowly precessing in the disk and creating the V/R cycles \citep{1991PASJ...43...75O,1992A&A...265L..45P}. 
The latter authors calculate that the radial extension of such a disk oscillation mode should be confined to a few stellar radii, which can be probed by interferometry.
Central properties of this theory have been observed indirectly by the continuous observation of individual line profiles \citep[one-armed prograde precession,][]{1994A&A...288..558T}, and the detection of phase-shifts between different emission line cycles pointing towards a spiral density perturbation \citep[][]{2007ApJ...656L..21W}. Spectro-interferometric observations of the photocenter shift in hydrogen lines of $\zeta$~Tau appear to directly confirm a one-armed density mode in the disk \citep{1998A&A...335..261V,2009A&A...504..915C,2009A&A...504..929S}.

After summarizing the observation and data reduction (Sect.~\ref{sec:2}), we will discuss in Sect.~\ref{sec:3} that our data show that the radial dependence of the hydrogen emission regions can be observed {\it directly} with KI-SPR, thanks to resolving several emission lines of the Pfund series together with \brg at unprecedented differential precision in the $K$-band. Concluding remarks are given in Sect.~\ref{sec:5}.

\section{Observations}
\label{sec:2}


\begin{table}
\begin{center}
\caption{\label{tab:11}Observing log from April 25, 2008 (UT). }
\begin{tabular}{ccccc}
\tableline\tableline
Name       & $V/H/K$$\,^{(a)}$ & UT & BL$_{\rm proj}$&  Stellar diam     \\
           &                   &    &                &  [mas] $\,^{(b)}$ \\
\tableline
\object{48 Lib} (Tar, B4III$\,^{(c)}$)    & 4.9/4.8/4.6 & 13:20:49 & 66.6~m, 39$^\circ$ EoN & $0.23\,\pm\,0.1$ \\
\object{HD 145607} (Cal, A4V) & 5.4/5.3/5.1 & 13:28:17 & & $0.25\,\pm\,0.04$ \\
\tableline
\end{tabular}
\tablenotetext{a}{From {\sc Simbad}}
\tablenotetext{b}{Bolometric photosphere fit using NExScI's getCal tool. Diameters $<$~0.5~mas are unresolved by the KI.}
\tablenotetext{c}{From \citet{1982ApJS...50...55S}. However the spectral type is uncertain by a few sub-classes as discussed for instance by \citet{1996A&A...310..849F}.}
\end{center}
\end{table}

Details of the observations are given in Table~\ref{tab:11}.
Each dataset consists of 155 frames of 0.5~sec integration time, taken on the second fringe camera. The frames were selected to have an absolute group delay of less then 4.5~$\mu$m to ensure high SNR and limited impact of the dispersion introduced by the air-filled delay lines \citep[see Appendix C in ][]{1999ApJ...510..505C}. The remaining 137 (151) frames of target (calibrator) were corrected for group delay (linear) and dispersion (quadratic) wavenumber slopes. The resulting pre-processed frames were stacked by calculating the mean in each wavelength bin, and the standard deviation of this mean is used throughout the paper as error. This statistical error is adequately describing the differential uncertainty of the data. The total flux data were flux normalized before the averaging. This is necessary to avoid that the larger absolute flux variation due to the spatial filtering of the single-mode fibre fed fringe camera dominates the here interesting differential flux variation between adjacent lambda bins. 

The resulting measurables of the SPR observation (flux, $V^2$, and \dph) are shown in Fig.~\ref{fig:1}. 
To achieve the rest wavelength calibration shown in the figure, two steps were performed. First, the observed wavelength calibration needs to be rectified. The slope is measured internally by a Fourier transform based technique. In standard KI observing setup, this slope is applied automatically to the data provided through the NExScI archive. Furthermore, we used telluric absorption lines in the raw spectra to estimate an offset in the wavelength calibration (typically of order of a few tenths of one spectral pixel), and to verify the archived rectification. The fact that our data do not show the full $K'$ band is due to an alignment problem of the early instrument setup used. For current performance, we refer to the KI support webpage \footnote{{\tt http://nexsci.caltech.edu/software/KISupport/v2/v2sensitivity.html}}. 
In a second step, the corrections for barycentric motion of the earth (-13~\kms) and peculiar motion of the star \citep[-6~\kms,][]{1967IAUS...30...57E} have been applied. 
Note for \brg, that the offset between zero \dph and rest wavelength is significantly larger than the accuracy of our wavelength calibration. It is attributed to observing a profile which is significantly asymmetric at the level of the used spectral resolution.
The quality of the resulting wavelength calibration is shown by overplotting the HI-recombination line centers in Fig.~\ref{fig:1} (dotted lines). It is reassuring, that (1) the flux line profiles are centered, that (2) the various Pfund lines show the same line profile, and most importantly, that (3) the \dph of both \brg and Pfund lines crosses zero close to the transition wavelengths, and with the {\it same} sign of the slope. (3) is expected for an inclined Keplerian disk.  

The target data were calibrated with a subsequently measured unresolved star, observed only 8~min later under equal atmospheric conditions. 
We applied simple calibration strategies: the raw target flux was divided by the calibrator, and multiplied by a measured spectral template from the IRTF spectral library (A3V, Vacca, personal comm.) containing the HI absorption features of the calibrator of similar spectral type. The spectral template provides a relative flux precision of 1~\%, and was folded to $R\,=\,1000$, and divided by a black body continuum fit to the line free regions before application to our data. The resultant spectrum (upper panel in Fig.~\ref{fig:1}) thus shows in each spectral bin the flux with respect to the calibrator.
The $V^2$ was corrected for the system visibility, estimated by observing an unresolved calibrator. 
We did not apply any flux bias correction in our data calibration. 
Although the primary fringe tracker at KI has a flux bias of a few percent per magnitude \footnote{{\tt http://nexsci.caltech.edu/software/KISupport/dataMemos/fluxbias.pdf}}
internal tests have demonstrated that the detector used for these science observations does not display such a bias.
The differential phase is calibrated by subtraction of the calibrator \dph.
To show the impressive differential stability of the data over the used band, we overplotted in red linear continua, fitted to the line-free regions ($\lambda\,\le\,2.15\,\mu$m and $2.18\,\le\,\lambda\,\le\,2.315\,\mu$m), over flux ratio and $V^2$. The slightly red slope of the linear flux ratio continuum probably originates in the shell-induced NIR excess of 48~Lib, which photosphere is hotter than the calibrator of spectral type A4V. The continuum trend in the visibility is due to resolving the innermost circumstellar gas, and is discussed in Sect.~\ref{sec:31}. The red line in the \dph panel marks zero phase, and is not a fit. The resulting pixel-to-pixel errors are $\Delta {\rm flux ratio}\,=\,0.003$, $\Delta\,V^2\,=\,0.007$, and $\Delta\,d\phi\,=\,3\,{\rm mrad}$. The higher relative precision of the \dph signal with respect to the precision of the visibility is expected for such spectro-interferometric measurements \citep{2006dies.conf..291M}.

This differential stability at the level of a few $10^{-3}$ is unprecedented for spectrally dispersed OLBI data at the spectral resolution and sensitivity offered by KI-SPR. It is the key to derive the scientific results presented in the next section. 
A more comprehensive description of the SPR mode and its sensitivity, performance, and technology will be given elsewhere \citep{2010Woi}. 

\section{Results}
\label{sec:3}

\begin{figure}
\includegraphics[angle=0,scale=.90]{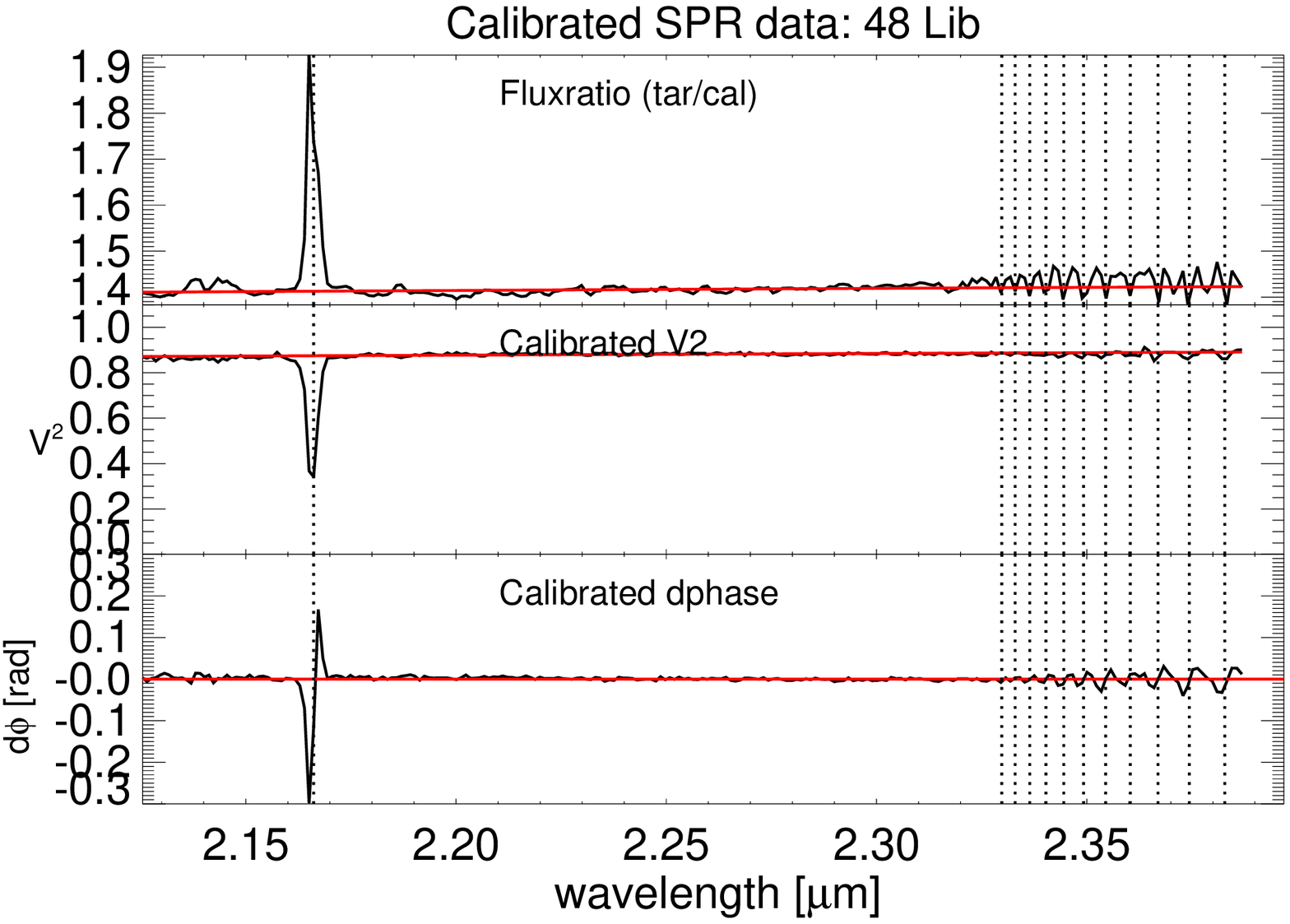}
\caption{{\it Top:} Mean calibrated flux ratio between the 48~Lib and the continuum divided calibrator.  The red solid line marks a linear continuum fit. The different line profiles of \brg and $Pf$-emission lines are clearly visible {\it Center:} Mean calibrated $V^2$ of 48~Lib showing that both the NIR continuum and the recombination line emission are spatially resolved by the interferometer. A linear continuum (red line) was fitted to the line free regions. {\it Bottom:} The calibrated differential phase data. The red line marks zero phase. All plots show the rest wavelength of the target. The vertical dotted lines indicate the rest wavelength of the recombination lines. It is apparent that all lines show the same slope at the line center, as expected for disk emission. \label{fig:1}}
\end{figure}

\begin{table}
\begin{center}
\caption{\label{tab:2}Measured and derived properties of the circumstellar shell of 48~Lib. \tablenotemark{a}}
\begin{tabular}{lccc}
\tableline\tableline
$K'_{\rm cont}$ (Gauss-disk only): &$\theta_{\rm cont}^{\rm FWHM}$& 0.94$\pm$0.03 mas&\\
$K'_{\rm cont}$ (Gauss-disk + star): &$\theta_{\rm cont,disk}^{\rm FWHM}$& 1.65$\pm$0.05 mas&\\
\tableline
\tableline
$Br\,\gamma$ & $V_{(-250\,,\,-100\,{\rm km\,s^{-1})}}$ & $R_{(100\,,\,250\,{\rm km\,s^{-1})}}$& $V\,/\,R$ \\
\tableline
$F_{\rm line}$ (norm.)                 & $0.29 \pm 0.01$ & $0.16 \pm 0.01$ & $1.80 \pm 0.2$ \\
$F_{\rm line}^{\rm corr}$ (norm.)      & $0.23 \pm 0.01$ & $0.14 \pm 0.01$ & $1.66 \pm 0.2$ \\
$\theta_{\rm line}^{\rm FWHM}$ (mas)   & $1.7  \pm 0.2$  & $1.4  \pm 0.2$  & $1.2  \pm 0.2$ \\
\pcs (mas)                             & $2.2  \pm 0.2$  & $1.9  \pm 0.2$  & $1.2  \pm 0.2$ \\
\tableline
\tableline
\pfa$\,(5:24..28)$ & $V_{(-325\,,\,-175\,{\rm km\,s^{-1})}}$ & $R_{(175\,,\,325\,{\rm km\,s^{-1})}}$& $V\,/\,R$ \\
\tableline
$F_{\rm line}$ (norm.)               & $0.029 \pm 0.01$  & $0.028 \pm 0.01$ & $1.0\pm0.1$ \\
$F_{\rm line}^{\rm corr}$ (norm.)    & $0.024 \pm 0.01$  & $0.024 \pm 0.01$ & $1.0\pm0.1$ \\
$\theta_{\rm line}^{\rm FWHM}$ (mas) & $1.9   \pm 0.3$    & $1.5   \pm 0.2$   & $1.2\pm0.3$ \\
\pcs (mas)                           & $0.9   \pm 0.3$    & $0.9   \pm 0.3$   & $1.1\pm0.3$ \\
\tableline
\tablenotetext{a}{All uncertainties given in the table are derived from the standard deviations of the mean total flux and correlated flux spectra shown in the figures. These uncertainties reflect the differential, {\it relative} precision of the values, and is adequately used throughout the article for intercomparison of the different components. However, the {\it absolute} accuracy of the individual estimates is $10\%$ for continuum and \brg values, and $25\%$ for the \pfa values.} 
\end{tabular}
\end{center}
\end{table}

Fig.~\ref{fig:1} shows that different components have been detected in the spectra of \object{48~Lib}: a clearly resolved continuum emission, and HI recombination line emission of \brg and various Pfund lines. 
The visibility data suggest that the continuum and the line emission stem from spatially separated, distinct regions in the circumstellar disk. 
Also, they derive from different emission processes. 
Therefore, we discuss in the following the two components separately.

\subsection{Continuum emission}
\label{sec:31}

The disk continuum emission, emitted by bound-free and free-free thermal emission in the disk, is clearly resolved. To facilitate the comparison of our results with other published disk diameters, we give the size of a disk-only fit to the continuum visibilities ($\theta_{\rm cont}^{\rm FWHM}\,=\,0.95\,\pm\,0.03~\,{\rm mas}$) in Table~\ref{tab:2}. We model the disk continuum by a maximally inclined Gaussian profile, with a position angle of 50$^\circ$ which is orthogonal to the axis of polarization \citep{1999PASP..111..494M}. However the measured continuum emission consists of both stellar and disk contributions. Taking into account, that the disk continuum of 48~Lib contributes one third of the $K$-band emission \citep{1994A&A...290..609D}, the resulting disk size scale ($\theta_{\rm cont, disk}^{\rm FWHM}\,=\,1.65\,\pm\,0.05~\,{\rm mas}$) is seven times larger than the stellar photospheric diameter. 
Source variability may compromise this estimation of $\theta_{\rm cont, disk}^{\rm FWHM}$. 
But a flux calibration of the KI photometry of 48~Lib, based on HD~145607 is consistent with the {\sc 2Mass} point source flux of our target \citep{2006AJ....131.1163S} to better than 10\%, which is the precision level of our absolute flux calibration. \citet{2009A&A...507.1141L} report an irregular variability in the Hipparcos band at the  5~\% level. Therefore, we can expect that variability of the star-disk flux ratio has no strong effect on the $\theta_{\rm cont, disk}^{\rm FWHM}$ value reported above.
Note that although important for the estimation of the disk size, we cannot derive the NIR disk excess from the visibility data due to our small $u,v$-coverage. Similarly, two-dimensional physical or phenomenological disk models of the continuum emission are not well constraint by a single baseline measurement.  
\citet{2007ApJ...654..527G} showed for several Be stars, that, even for significantly two-dimensional $u,v$-coverages, physical models, using radiative transfer calculations, and phenomenological models, using Gaussian intensity profiles are difficult to distinguish by the visibility data alone.
Physical and phenomenological FWHM estimates of the average disk compactness may differ by 50-100~\% \citep{2007ApJ...654..527G}.
This may be used as an accuracy estimate for a physical interpretation of the here derived $\theta_{\rm cont, disk}^{\rm FWHM}$. Since near-infrared disk excess and $\theta_{\rm cont, disk}^{\rm FWHM}$ of 48~Lib are within the typical range of Be star disks, we can expect that 
a physical modeling of the disk continuum, as done by \citet{2007ApJ...654..527G}, would lead to similar densities as described therein.

The estimated disk continuum-to-stellar radii ratio matches other interferometrically observed edge-on systems with comparable properties \citep[e.g. $\zeta$ Tau in ][]{2007ApJ...654..527G}. 
The typically found moderate extent of the $K$-band emission size (5-10 times the stellar radius) is significantly smaller than recently modeled \citep{2001A&A...367..532S}, confirming the need for interferometric constraints to understand the physical conditions in the stellar outflows of Be-stars.

There is several indications in our data, that the disk continuum emission zone is inside the \brg and smaller or equal to the $Pf-$ recombination line emission zones.
All fitted linear characteristics of the violet and red part of the \brg (\pfa) line emission zones (that is $\theta_{\rm line}^{\rm FWHM}$ and the photocenter shifts \pcs, for details see the next section; violet and red labels negative and positve Doppler shifts, following standard Be-star terminology) indicate that the majority of the line emission is created at stellocentric radii larger than (comparable to) the radii dominating the continuum emission. 
Furthermore, the \pfa line center in total and correlated flux drops below the continuum level (Fig.~\ref{fig:3}). This points to a significant absorption of the continuum flux by gas {\it in front} of the continuum source, that is the continuum emission is embedded and surrounded by the zone dominating the line emission. 
Therefore, we confirm the findings of \citet[][studied $H\,\alpha$ versus continuum size]{2007ApJ...654..527G} and \citet{2009A&A...504..929S} that the observed \brg ~(and \pfa) hydrogen line emission in Be stellar shells appears to occur at larger (or equal) stellocentric radii than the disk continuum emission.

\subsection{HI recombination lines}
\label{sec:32}

\begin{figure}
\includegraphics[angle=0,scale=.90]{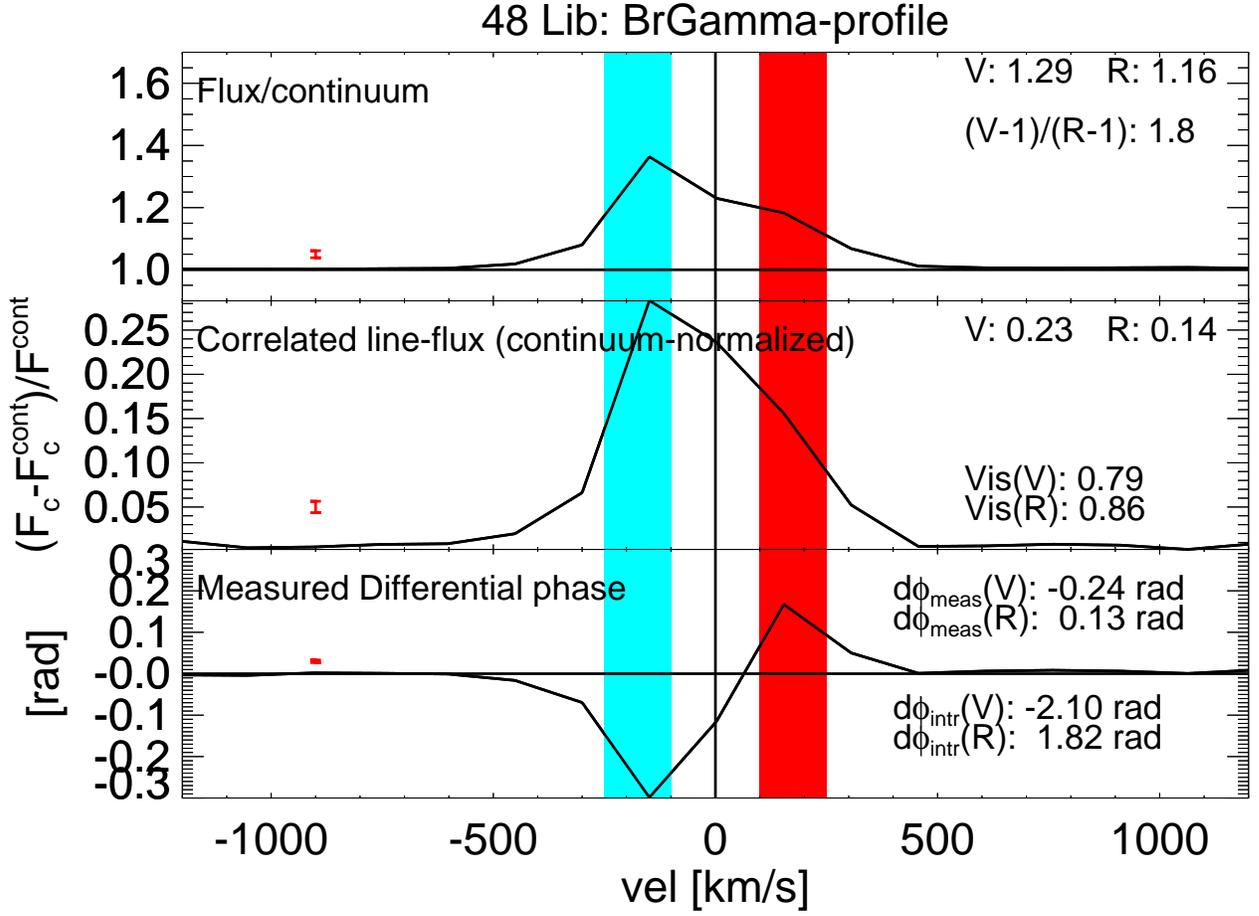}
\caption{Zoom into the \brg line. The flux ratio ({\it top}) is normalized to the linear continuum shown in Fig.~\ref{fig:1}. The {\it central} panel shows the correlated flux, normalized to the linear continuum fit. 
Note that the linear flux ratio continuum fit may overestimate the actually observed continuum emission 
due to absorption by gas {\it in front} of the continuum emission, which would result in underestimated fluxes at the line center. 
The measured differential phase is plotted in the {\it bottom} panel. 
At the top right of each figure, the integrated violet and red line profile is given. The red area marks the region of integration (100-250~km/s), which excludes the uncertain central continuum absorption. The wavelength-differential precision is indicated by the red errorbar in the left of each panel. \label{fig:2}
}
\end{figure}

\begin{figure}
\includegraphics[angle=0,scale=.90]{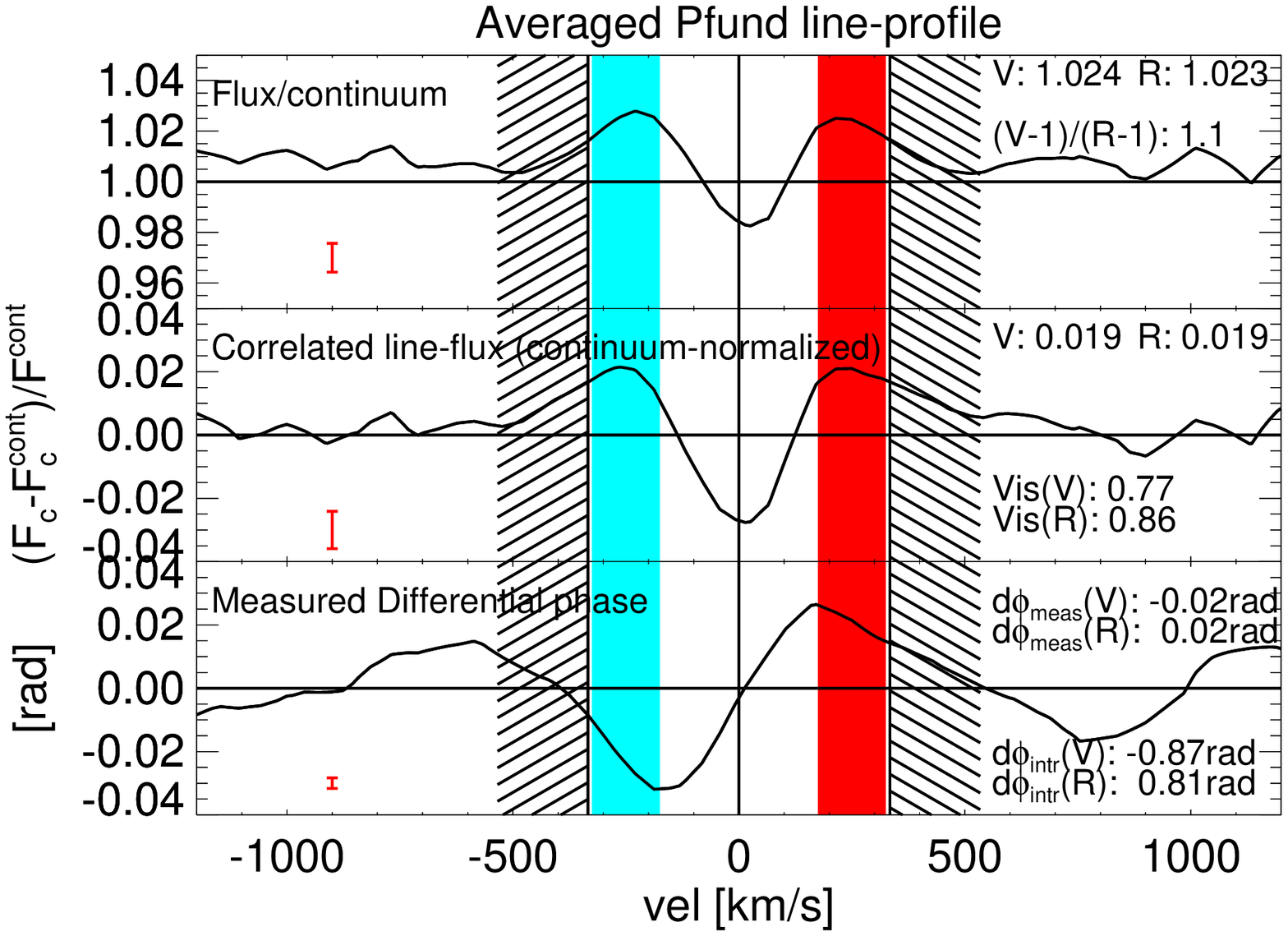}
\caption{The three panels show the \pfa profile at the flux levels of Pf-(5-24) in the same way as in Fig.~\ref{fig:2}. The smoother line profile with respect to \brg derives from the line averaging over the linearly interpolated individual Pfund lines. The hatched area indicates the beginning of the velocity region where the shape of the mean line profile is significantly compromised by the neighboring lines due to the averaging. The profile at large velocities ($\ge\,330\,{\rm km\,s^{-1}}$) are a result of our data processing, and unreal. We show this part here to demonstrate that the central, real line profile is more significant than these relatively large profile changes in the outer velocity-space compromised from the averaging. The statistical error of the mean shapes shown is indicate by the red errorbar in the left of each figure. \label{fig:3}}
\end{figure}

The emission line widths of Be-stars are dominated by large-scale kinematic Doppler broadening (due to the fact that the ionized gas is located in a rotating disk around the star), without significant contributions of non-kinematic scattering processes \citep{1989Ap&SS.161...61H,2007ApJ...654..527G}.
The disk of 48~Lib is known to be edge on, which simplifies the geometric and kinematic interpretation of the line profiles \citep{1996A&A...312L..17H}. 
The spectrometer does not fully Nyquist-sample the double peak of the \brg and Pfund lines, but the central dip in the flux and the shape of the \dph~signal show that the violet (V) and red (R) part of the disk are clearly separated by the spectrometer, and that meaningful V/R ratios can be derived. 

$Pf$-24..34 have been detected for the first time in an interferometric spectrum. The five strongest individual $Pf$-lines are clear detections ($\sim\,5-8\,\sigma$) above the differential precision level of flux ratio, visibility, and \dph. To improve on the SNR for the $Pf-$lines, and to minimize the effects of the limited sampling of the line-profile, we average over the five strongest, detected $Pf-$lines (24..28). To do so, the $Pf-$25..28 lines were scaled to match flux levels of $Pf-$24. This avoids that the final average profile (\pfa) is affected by the individual line fluxes. We checked that none of the  features of \pfa discussed below is significantly altered or vanishes when averaging over all 11 detected Pfund lines. However, our choice of averaging over the five strongest lines to create \pfa leads to the highest SNR. 
In addition, the omitted higher-order Pfund lines are closer in rest wavelength, and therefore, overlapping of adjacent line profiles would become apparent and compromise the profile of \pfa. Similarly, at velocities $\ge 330~$\kms (indicated by the dashed region in Fig.~\ref{fig:3}), the average profile is affected by adjacent lines. We purposefully show in the figure a large velocity space including these artifacts to demonstrate that \pfa emerges even out of these averaging artifacts. 
Unless indicated otherwise, we use a linear flux ratio continuum fit. Using a black body instead of a linear shape for the flux ratio continuum leads to a comparably good fit, and slightly increases the off-center line emission of \pfa. We include this uncertainty in the shape of the continuum at the red end of the K'-band in the errors given in Table~\ref{tab:2}. 
The statistical precision of the averaged profiles are shown as red errorbars.

We compared the individual properties (flux, visibility, and differential phase) of the $Pf-$(24..28) included in \pfa to check if meaningful geometric constraints on the Pfund line emission zone can be derived from \pfa . The decreasing signature in visibility and phase with increasing excitation level is consistent with the $Pf$-n/$Pf-$24 line flux ratios. This means that a similar geometry is encoded in the measured visibility and phase of each line, suggesting a common origin of the lines, and justifying the averaging.
The individual line ratios used match the respective Pfund-line ratio trends measured for other Be-star envelopes, such as \object{$\gamma$~Cas} \citep{2000A&A...355..187H} and \object{BK~Cam} \citep{2010AJ....139.1983G}. Furthermore, these flux ratios vary by no more than up to about 30~\% ($Pf$-28/$Pf-$24), which indicates together with a similarly moderate variation of the respective Einstein coefficients that the lines included in \pfa are emitted under similar conditions. Therefore, they most likely originate from similar regions in the disk. The geometric constraints, derived from \pfa, cannot be more accurate than this level, which matches the numbers, given in Table~\ref{tab:2}.

The precision of the $V/R$-estimate of the individual Pfund lines is clearly limited by our spectral resolution and the sampling of the line profile. Apparent ratios of up to 20~\% above and below unity for the Pfund lines included in \pfa are regarded as not significantly different from unity. Therefore, the lines show comparable $V/R$. Given the previous arguments for a common origin of the individual lines in the disk, we can expected that averaging does improve the $V/R$-signal for the Pfund emission zone.

We concentrate on answering the key question qualitatively: are KI-SPR data suitable to measure directly, and efficiently in a single dataset, core parameters like radial extent and radius-dependence of the often discussed slowly rotating density inhomogeneity in the gas around Be stars?
To do so, violet and red velocity bins outside of the line center were defined to derive average V-R properties of the lines, while avoiding the zone of highest absorption and uncertain continuum (see below). The velocity bins enclose the maximum line emission with a width of 150~\kms, which is chosen to be comparable to the spectral resolution (bins are marked as colored background in Fig.~\ref{fig:2}\&\ref{fig:3}). 
The profile averages for each bin are given in the figures. We derived intrinsic line emission sizes and photo-center shifts from the data (Table~\ref{tab:2}) by assuming the superposition of a linear continuum and additional \brg disk emission \citep[for the derivation of the relevant formulae see Appendix C in][]{2007A&A...464...87W}. Note that the intrinsic photocenter shifts are on the order of the fringe spacing ($\lambda /2B$), which requires the use of the exact complex relation between measured and intrinsic phase \citep[Eq.~C.3 in][]{2007A&A...464...87W} to derive the correct \pcs.  Table~\ref{tab:2} reports the mean linear properties fitted to the linearly interpolated data for each velocity bin. 

For our analysis, we used a linear continuum extrapolation of the continuum flux ratio into the line center to derive the quantitative constraints in Table~\ref{tab:2}. However, the actual continuum at the \brg line center may deviate from this linear continuum.
The stellar continuum has some photospheric \brg absorption, and since we observe an edge-on disk in emission with, at least partially, optically thick \brg emission, additional absorption and emission makes it difficult to estimate the exact continuum shape at the line center.
This presents a difficulty for isolating the correlated flux, visibility and intrinsic \dph of the line emission at the line center from the measured data, which contain a flux-weighted average of both continuum and line emission. 
An absorbed continuum profile (in contrast to the here assumed linear extrapolation of the off-line continuum) at the center of \brg would increase the (correlated) line flux, and therefore decrease the derived intrinsic photocenter shift. To quantify this, we also applied a B4III \brg template to the data (see also footnote $^{\rm (c)}$ in Table~\ref{tab:11}). This would not change the measured V/R asymmetries significantly, but it would increase the given \thl by about 40~\% and decrease \pcs by 20\%, which gives the order of magnitude of this systematic uncertainty in our quantitative analysis.

\subsection{Radial disk structure\label{sec:33}}

\begin{figure}
\includegraphics[angle=0,scale=.70]{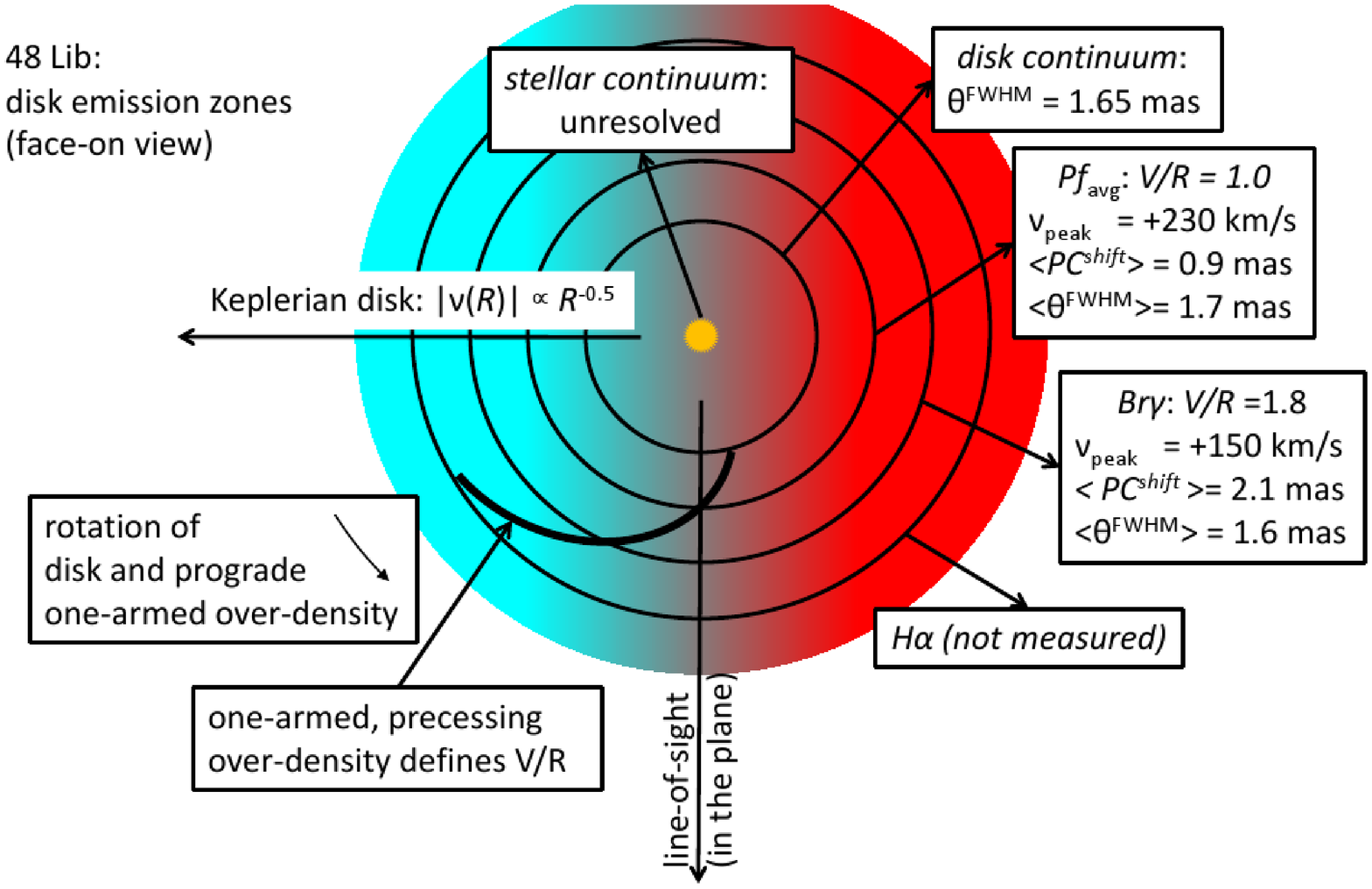}
\caption{Visualization of the measured disk properties (not to scale). The shown one-armed over-density pattern would explain the measured {\it V/R} and correlated flux profiles of \brg and \pfa, as predicted by \citet{1992A&A...265L..45P}, but the here shown pattern is not based on a model calculation. The exact shape of the pattern is not constrained by a single KI-SPR dataset. The relative location of the optical $H\alpha$ line is added for completeness, matching previous, single telescope velocity and {\it V/R} measurements \citep{2007ApJ...656L..21W}.  \label{fig:4}}
\end{figure}

\brg  and \pfa profiles differ. Above all, they show different V/R ratios and central absorption levels. The total and correlated \brg line profiles are with $V/R\,\sim\,1.8$, and 1.6 respectively, significantly asymmetric. In contrast, \thl and \pcs of the violet and red \brg emitting regions coincide within the uncertainties. 
\citet{2009A&A...504..929S} resolved the \brg disk emission of $\zeta$~Tau with the VLTI with a spectral and spatial resolution similar to our data. They found a larger intrinsic line photocenter shift of the brighter wing, emitted, at the time of their observation, from the south-eastern, red part of the disk of $\zeta$~Tau. A non-LTE disk model can explain such a trend in the \dph as well as an extensive multi-wavelength and multi-technique dataset with a global oscillation of a spiral density pattern \citep{2009A&A...504..915C}.
Our data might show a similar trend of a slightly larger \pcs of the brighter side of \brg. However, this trend is not significant with respect to the estimated uncertainties. We would encourage a similar, complex modeling effort of the disk of 48~Lib, including the new differential constraints from the Pfund emission and based on repeated observations to improve on the significance of the \dph trend of \brg. Here we concentrate on the overall line emission zone properties.

The bulk of the \brg emission appears to come from similar stello-centric radii, and the enhanced emission on the violet side indicates a locally enhanced  density.
The observed asymmetry is inline with the $\sim$9~yr periodicity of cyclic changes of V/R($H\,\alpha\,$)  \citep{1995A&A...300..163H}. 
The $H\,\alpha$ spectra of 48~Lib, provided by the {\sc BeSS} database \footnote{{\tt http://basebe.obspm.fr/basebe/}} also report a V/R $>\,1$ in 2008.
In contrast, the \pfa profile appears to be symmetric in all properties within the errors. Again, \thl and \pcs of \pfa are comparable for both sides, but the smaller \pcs points to significantly smaller stello-centric radii of the bulk of the \pfa emission, if compared to \brg. Both measured V/R(\brg) and V/R(\pfa\nolinebreak) can be reconciled with a one-armed density perturbation, if a radial dependence of the perturbation is allowed, as in the case of a spiral density wave, precessing through the disk. 

There is several indications for that the Pfund emission emerges from inside the bulk of the \brg emission. 
The mean \pcs(\brg) is with 2.1~mas about twice as large as \pcs(\pfa\nolinebreak).
Furthermore, the bins are centered at different stello-centric radii and velocities.
The peak velocities appear offset (\brg: $150\,\pm\,50\,$\kms versus \pfa: $230\,\pm\,30\,$\kms). 
These velocity and \pcs numbers are consistent with Keplerian gas motion, a typical assumption for Be star disk. The Keplerian rotation assumption is further supported by the measured double-peaked line profiles (in contrast to highly disordered or asymmetric profiles).

Since Be-star disks are (at least partially) optically thick in the hydrogen emission lines \citep[see the central absorption in Fig.~\ref{fig:3} and the Be-star disk study of ][]{2002A&A...386L...5L}, it is expected that the stellocentric radius of the respective line anti-correlates with the specific line absorption properties. Comparing the respective Einstein absorption coefficients, we find $B_{5,24}\,<\,B_{4,7}\,<\,B_{2,3}$, which suggests that the \pfa is emitted inside of the \brg as described by our data. \brg should be emitted at smaller stellocentric radii than $H_{\alpha}$, as indirectly confirmed by the data of \citet{2007ApJ...656L..21W} \citep[similarly for $\zeta\,$Tau:][]{2009A&A...504..915C}. Also the smaller $H_{\alpha}$ peak velocities \citep[the spectrum in Fig.~1 of][suggests about 110 km /s]{2007ApJ...656L..21W} are consistent with larger stellocentric radii compared to \brg and \pfa.

Another indication for the different location of the Pfund line emission are the total and correlated flux profiles. Both show a clear absorption below the continuum emission level. This, together with the symmetric line profile might indicate that the gas dominating the Pfund emission is located {\it in front} of the inner disk and the star (responsible for the continuum) \citep[see the extensive line profile modeling of ][]{1994IAUS..162..382H, 1994A&A...288..558T}. As discussed, we cannot estimate what fraction of this continuum absorption is due to intrinsic photospheric absorption. But a domination of the measured continuum absorption by photospheric absorption is unlikely for a Be star. Having the majority of the \brg emission at larger stellocentric radii on the violet side, but the Pfund emission at smaller radii in front of the star suggests that the over-density pattern follows a radial, one-armed spiral pattern  \citep{2007ApJ...656L..21W,2009A&A...504..915C}.
We depict the radial structure of the disk emission as measured by the KI in a schematic way in Fig.~\ref{fig:4}. However, our single dataset obviously cannot distingiush a spiral wave from other radius-dependent perturbations patterns.

Our data provide a further quantitative test of Papaloizou's density wave model. \citet{1992A&A...265L..45P} discuss that the temporal prediction of V/R cycles of about 10~yr periods fits the available spectroscopic observations. 
Interferometric observations directly resolve the expected linear scale of the predicted density waves, as opposed to indirect methods such as polarimetry \citep[as discussed for instance in ][]{2010ApJ...709.1306W}.
The mean photocenter shifts result in stellocentric radii of about 18~(8)~$R_*$ for \brg (\pfa), which is consistent with both hydrogen emission line modeling of Be stars \citep[e.g.][]{2000A&A...359.1075H,2000A&A...355..187H} and with density wave models predicting that the local density perturbation, induced by the rotational flattening of the Be stars, disappears outside a few stellar radii \citep{1992A&A...265L..45P}.

\section{Conclusions}
\label{sec:5}

We presented the first interferometric detection of several Pfund lines at the red end of the $K'$-band.
Thanks to the simultaneous detection of \brg in the spectrum of a single KI-SPR observation, we can derive stello-centric radius dependent properties the gas disk of the classical Be-star 48~Lib at high precision.
The geometry and kinematics of the emission zones of \brg and of eleven higher Pfund lines  are clearly detected in the differential visibility and phase signals, and the Pfund emission originates from smaller stello-centric radii than \brg.
The radial separation of the various emission lines is convincingly explained by the physics of optically thick line emission in the Be circumstellar disks. 
In addition, the $K'$-band disk continuum emission was resolved as well, and shown to be emitted from inside the \brg (V/R $>1$) hydrogen line zone, at radii comparable to the bulk of the $Pf$-emission (with V/R $\approx\,1$).
Our findings match qualitatively and quantitatively theoretical Be-star disk model calculations, predicting a precessing one-armed over-density pattern.
The spatial constraints from our interferometric data coincide with the common notion that Be-star disks have a Keplerian-like radial velocity profile.
Observations similar to the here presented will shed more light on the enigmas of the creation and properties of circumstellar Be shells at all stages of evolution. In particular time-resolved monitoring of Be circumstellar environments with KI-SPR together with radiative transfer modeling of the disk emission will deliver strong constraints on the actual radial and azimuthal geometry of the disk inhomogeneities likely to be responsible for the HI line emission asymmetries.

In addition to the presented scientific results on the disk of 48~Lib, we demonstrate with the detection of the Pfund lines, that even weak line emission can produce a significant signal in the differential phase.
KI-SPR data of bright stars can be calibrated to a differential precision of $\Delta {\rm flux ratio}\,=\,0.003$, $\Delta\,V^2\,=\,0.007$, and $\Delta\,d\phi\,=\,3\,{\rm mrad}$.
This extreme precision in the differential phase equals an on-sky centroid shift of 3~$\mu$as (a relative precision of 10$^{-3}$), which is an improvement by two orders of magnitude over state-of-the-art single-telescope spectro-astrometry. This improvement is however limited to the relatively small interferometric field of view ($\sim$30 mas) and sensitivity of the current KI-SPR mode. 
Dual field phase referencing, the second phase of the KI-ASTRA upgrade, is currently under construction, and will allow fainter science targets to be observed. 
Using a different dispersive element, even higher spectral resolution could be achieved in SPR mode. 



\acknowledgments

We are grateful to W. Vacca and A. Seifahrt for many useful discussions and contributions to this work. The excellent support of the KI team at WMKO and NExScI helped making these observations a success.
The realization of the KI-ASTRA upgrade is supported by the NSF MRI grant, AST-0619965.
The data presented herein were obtained at the W.M. Keck Observatory, which is operated as a scientific partnership among the California Institute of Technology, the University of California and the National Aeronautics and Space Administration. The Observatory was made possible by the generous financial support of the W.M. Keck Foundation. The authors wish to recognize and acknowledge the very significant cultural role and reverence that the summit of Mauna Kea has always had within the indigenous Hawaiian community.  We are most fortunate to have the opportunity to conduct observations from this mountain. The Keck Interferometer is funded by the National Aeronautics and Space Administration as part of its Exoplanet Exploration program. This research has made use of the SIMBAD database,
operated at CDS, Strasbourg, France. This research has made use of NASA's Astrophysics Data System Bibliographic Services. This work has made use of the BeSS database, operated at GEPI, Observatoire de Meudon, France.


{\it Facilities:} \facility{Keck:I ()}, \facility{Keck:II ()}

%

\end{document}